\RequirePackage{fix-cm}
\documentclass[smallextended]{svjour3}       % onecolumn (second format)
\smartqed  % flush right qed marks, e.g. at end of proof
\newenvironment{numberedlist}
{\begin{list}{\makebox[20pt]{\hss(\arabic{itemno})\enspace}}
             {\usecounter{itemno}\labelwidth 20pt}}{\end{list}}

\newcounter{itemno}

\newcounter{itemno1}

\newcounter{itemno2}
\newcounter{exno}

\newcounter{defno}

\newenvironment{defn}{\refstepcounter{defno}\medskip \noindent {\bf
Definition \thedefno.\ }}{\medskip}

\newcommand{\sep}{\;\vert\;}

\newcommand{\oprove}{\vdash\kern-.6em\lower.7ex\hbox{$\scriptstyle O$}\,}

\newcommand{\Pscr}{{\cal P}}

\newcommand{\pderivation}{{\cal P}\kern -.1em\hbox{\rm -derivation}}
\newcommand{\pderivationl}{{\cal P}\kern -.1em\hbox{\em -derivation}}
\newcommand{\pderivable}{{\cal P}\kern -.1em\hbox{\rm -derivable}}
\newcommand{\pderivablel}{{\cal P}\kern -.1em\hbox{\em -derivable}}
\newcommand{\pderivations}{{\cal P}\kern -.1em\hbox{\rm -derivations}}
\newcommand{\pderivability}{{\cal P}\kern -.1em\hbox{\rm -derivability}}

\newcommand{\all}{\forall}
\newcommand{\some}{\exists}

\newcommand{\ie}{{\em i.e.}}

\newsavebox{\lpartfig}
\newsavebox{\rpartfig}

\newenvironment{exmple}{
 \begingroup \begin{tabbing} \hspace{2em}\= \hspace{3em}\= \hspace{3em}\=
\hspace{3em}\= \hspace{3em}\= \hspace{3em}\= \kill}{
 \end{tabbing}\endgroup}

\newcommand{\lb}{\langle}
\newcommand{\rb}{\rangle}

\newcommand{\Ra}{\supset}  
\newcommand{\seqand}{\land^s}
\newcommand{\seqor}{\lor^s}

\newcommand{\seqweb}{{SeqWeb}}

\begin{document}

\title{Sequential Operations in LogicWeb%\thanks{Grants or other notes
}

\author{ Daeseong Kang  \and Keehang Kwon  \and Zulkarnine Mahmud
         %etc.
}


\institute{Daeseong Kang \and  Keehang Kwon \and Zulkarnine Mahmud \at
              School of Electronics and Computer Eng., DongA Univ, 604-714 Busan, Korea\\
             % Fax: +123-45-678910\\
              \email{khkwon@dau.ac.kr}           %  \\
 }


\maketitle

\begin{abstract}
Sequential tasks cannot be effectively handled in logic programming
based on classical logic or linear logic. This limitation can be addressed by
using a fragment of Japaridze's computability logic. We propose \seqweb, an extension to LogicWeb with sequential goal  
formulas.
SeqWeb extends the LogicWeb by allowing goals of 
the form $G\seqand G$ and $G\seqor G$  
 where  $G$ 
is a goal.   These goals allow us to specify both
sequential-conjunctive and sequential-disjunctive tasks.
\end{abstract}

\section{Introduction}
\label{intro}

\subsection{LogicWeb}

Logic languages such as Prolog support $nondeterminism$ (and automatic backtracking)
 and thus
is at a higher level of abstraction than imperative
programming languages.  Unfortunately, logic programming has been popular only
in limited applications including AI, expert systems, and compilers.
In particular, logic programming languages  has been little used  for Web programming. 
Only recently has the work called LogicWeb \cite{LD96} begun on the relationship between logic programming and
the Web. LogicWeb views the Web as a collection of
logic programs. 
   
LogicWeb thus allows us to build a web of information that machines
can understand and reason about. This includes the ability to answer
complex queries.  The standard language that has been introduced so far is a subset
 of first-order logic called Horn clauses with embedded implications.

    Implementing LogicWeb, however, requires more than Horn clauses with embedded implications.
It must be able to deal with name scoping or local constants to support information hiding.
       A language called first-order hereditary Harrop formulas 
($fohh$) \cite{Mil89jlp,MNPS91,Kwon08b} has been extensively studied. This language offers 
 modularity abstraction with local constants.

\subsection{First-order herediatary Harrop formulas}\label{sec:examples}
 
 The syntax of goals (or queries) permitted in Prolog is the
following: they can be atomic formulas or conjunctions or disjunctions 
of  goals. In the context of Prolog, conjunctions are written
using commas, while disjunctions use semicolons.

The language of $fohh$ extends
this set of logical symbols  to include implications and
universal quantifiers. In this language, formulas such as $D \supset
G$ and $\all x G$ will be permitted as goals, provided that $D$ is a program and $G$ is itself a
goal. The intended semantics of these two new operations is the
following. A goal such as $D \supset G$ is to be solved by adding $D$
to the current program.  Hence,
 $D$ is to be available only during the course of solving $G$. As for
a goal such as $\all x G$, it is intended to be solved by
instantiating $x$ in $G$ by a new constant $c$ and then solving the
resulting goal. Interpreted in this fashion, the universal quantifier provides
a means for limiting the availability of names. 

We illustrate the problem mentioned above by considering the definition of the 
fibonacci relation in Prolog. 
An  efficient fibonacci program can be written by using the idea of
an accumulator.  Implementing this idea in Prolog requires an auxiliary
predicate, called $fib\_aux$ below, to be defined. 
The following
definition is a traditional way of realizing this:
\begin{exmple}
\> $module\ fib$. \\
\> $fib(N,F)$ {\tt :-} $fib\_aux(2,N,1,1,F).$\\
\> $fib\_aux(M,N,F1,F2,F2)$ {\tt :-} $M >= N.$ \% Nth fibonacci number reached \\
\> $fib\_aux(M,N,F1,F2,F)$ {\tt :-} $M < N, fib\_aux(M+1,N,F2,F1+F2,F).$
\end{exmple}
\noindent The declarative interpretation of $fib\_aux$ here is
that it is true  if $F1$ is the $(M-1)$th, $F2$ is the $M$th fibonacci number
and $F$ is the $N$th fibonacci number.  In this example, it is
worth noting that $fib\_aux$ is a specialized
predicate whose only purpose for existence is its usefulness in
defining $fib$.

The addition of implication to goals provides a means for solving at
least part of this problem. Thus, the definition of $fib\_aux$ can
be made ``local'' to that of $fib$, as indicated below:
\begin{exmple}
\> $fib(N,F))$ {\tt :-}\\
\> $ (module\ fiba) \supset fib\_aux(2,N,1,1,F).$
\end{exmple}
\noindent where
\begin{exmple}
\> $module\ fiba$. \\
\> $fib\_aux(M,N,F1,F2,F2)$ {\tt :-} $M >= N.$ \% Nth fibonacci number reached \\
\> $fib\_aux(M,N,F1,F2,F)$ {\tt :-} $M < N, fib\_aux(M+1,N,F2,F1+F2,F).$
\end{exmple}

\noindent  Given this definition of $fib$, the
following points might be observed. First, the clauses defining $fib\_aux$
are not available at the top-level. 
Secondly, while the definition of  $fib\_aux$ is not available at the
top-level,  we see that it will become 
available when solving the body of $fib$. 

 However, there is one problem that still remains. The
meaning of the predicate $fib\_aux$ {\it inside} the body of $
fib$ is not insulated from definitions in existence outside the body.
This  problem may be solved by using a universal
quantifier. 
\begin{exmple}
\> $fib(N,F))$ {\tt :-}\\
\> $ \all fib\_aux\ ((module\ fib) \supset fib\_aux(2,N,1,1,F)).$
\end{exmple}
\noindent The indicated semantics of the universal quantifier dictates
picking a new name for $fib\_aux$ and then solving the
instantiation of the given query with this name.

The examples presented here have been of a very simple nature.
They are, however, sufficient for understanding the intended semantics
of the new logical symbols and also for appreciating some of their value from
a programming perspective. 
% More extensive exmples may be found in 
% various places in the literature (see, for example,
% \cite{Felty89phd,Mil89jlp,Mil89,Hannan90,PM90iclp}). 
We note that the usefulness
of the scoping mechanisms provided by these symbols have been extensively
studied and we point the interested reader to, e.g. 
\cite{Mil89jlp}.

\subsection{Contributions of this paper}

Although the class of $fohh$ is quite attractive, sequential tasks cannot be naturally handled in it.
This limitation can be addressed by
using a fragment of Japaridze's Computability Logic(CL) (\cite{Jap03} -- \cite{JapCL12}).  CL is a new semantic platform for reinterpreting logic
as a theory of computable tasks. Formulas in CL stand for instructions
that can carry out some computational tasks. Queries stand for actually
executing these instructions.

This paper proposes Sequential LogicWeb (\seqweb), an extension of LogicWeb \cite{Kwon08b} with sequentiality in goal formulas. Traditionally, LogicWeb has the problem of expressing sequential tasks.
 The class of sequential goal formulas enables the programmer
to express these sequential tasks. They offer the possibility to combine
imperative programming with declarative programming.

We need to define some terminology. A {\it move} (or a choice) is a string over the keyboard alphabet.
To be specific, executing $\some x G$ where $G$ is a goal requires a move by the machine. Similarly,
executing $\all x C$ where $C$ is a clause also requires a move by the machine.

\begin{itemize}

\item A sequential-multiplicative-conjunctive(SMC) goal  is of the form $G_1 \seqand G_2$ 
 where $G_1, G_2$ are goals.
Executing this goal has the following intended semantics: execute both $G_1$ and $G_2$ sequentially.
In other words, all the moves in $G_1$ must precede all the moves in $G_2$.
 Both executions must succeed for executing $G_1 \seqand G_2$ to succeed.
 It can be seen as a restricted version of the parallel-conjunctive goal $G_1 \land G_2$ where
all the moves in $G_1$ must precede all the moves in $G_2$

\item A sequential-multiplicative-disjunctive(SMD) goal is of the form $G_1 \seqor G_2$ 
 where $G_1, G_2$ are goals.
Executing this goal has the following intended semantics: execute both $G_1$ and $G_2$ sequentially. 
At least one of these two executions must succeed
for executing $G_1 \seqor G_2$ to succeed. A SMD goal can be used to express {\it success with a degree}.
That is, new successes are seeked, even after a success(solution) was found, as long as there probably are more successes to come.
Imagine Kim decides to buy a used BMW from the local car dealers. Everytime he finds one at a local 
dealer(\ie, a success), he will probably go to another dealer for a cheaper price(\ie, for another success).
Coming up with multiple successes does not affect the successfulness of the effort, but the behaviors of the subsequent tasks
such as how much Kim has to pay for the car
(Remember that we  now have the notion of sequential tasks, hence the notion of subsequent tasks.).

\end{itemize}

In this paper we present the syntax and semantics of this extended language, 
show some examples of its use and study the interactions among the newly added
constructs.  

The remainder of this paper is structured as follows. We describe \seqweb\
based on a first-order sequential hereditary Harrop formulas in
the next section. In Section \ref{sec:modules}, we
present some examples of \seqweb.
Section~\ref{sec:conc} concludes the paper.

\section{The Language}
\label{sec:logic}
The language we use is an expanded version of hereditary Harrop
formulas. It is described
by $G$- and $D$-formulas given by the syntax rules below:
\begin{exmple}
\>$G ::=$ \>  $A \sep G \land  G \sep G \lor G \sep D \Ra G \sep \all x\ G \sep \some x\ G \sep G\seqand G 
\sep G \seqor G $ \\    
\> \\   
\>$D ::=$ \>  $A  \sep G \supset A\ \sep \all x\ D  \sep  D \land D $\\
\end{exmple}
\noindent

In the rules above, $A$  represents an atomic formula.
A $D$-formula  is called a  
 clause or an instruction. 
 
In the transition system to be considered, $G$-formulas will function as 
queries and a set of $D$-formulas will constitute  a set of available
instructions. For this 
reason, we refer to a $G$-formula as a  a query, to a set of $D$-formula 
as an instruction set. Our language is an extension to first-order Hereditary Harrop formulas 
with the main difference that new sequential constructs are 
added in $G$-formulas. 

 We will  present an operational 
semantics for this language. These rules in fact depend on the top-level 
constructor in the expression and have the effect of producing a new 
program and a new available instruction set.

The rules for executing queries in our language are based on
``goal-directness''  in the sense that the next rule to be used depends on the 
top-level construct of the goal formula. This property is known as
uniform provability\cite{MNPS91}. This property guarantees  that the programmer knows exactly how execution will proceed.

\begin{defn}\label{def:semantics}
Let $G$ be a goal and let $\Pscr$ be a finite set of available instructions.
Then the notion of   executing $\lb \Pscr,G_1\rb$ -- executing $G$ relative to $\Pscr$ -- 
 is defined as follows:

\begin{numberedlist}

\item  If $G$ is an atom and is identical to an instance of a program clause in $\Pscr$,
 then   the
current execution terminates with a success. 

\item   If $G$ is an atom and an instance of a program clause in $\Pscr$ is
 of the form
 $G_1 \supset G$, execute  $\lb \Pscr, G_1\rb$. This execution must succeed for the
current execution to succeed.

\item If $G$ is $G_1 \land G_2$, then execute both $\lb \Pscr,G_1\rb$ and
$\lb \Pscr,G_2\rb$ in parallel. 
 Both  executions must succeed for  the current execution to succeed.

\item If $G$ is $G_1 \lor G_2$, then  execute both $\lb \Pscr,G_1\rb$ and
$\lb \Pscr,G_2\rb$ in parallel.  At least one of
these two  executions must succeed for the current execution to succeed.

\item If $G$ is $\exists x G_1$, then execute $\lb \Pscr,[t/x]G_1\rb$
where $t$ is a  term. This execution must succeed for the
current execution to succeed.

\item If $G$ is $\all x G_1$, then execute $\lb \Pscr,[a/x]G_1\rb$
where $a$ is a new constant.  This execution must succeed for the
current execution to succeed.

\item If $G$ is $D \Ra G_1$
, then execute $\lb \{ D \} \cup\Pscr, G_1\rb$. This execution must succeed for the
current execution to succeed.

\item If $G$ is $G_1\seqand G_2$, then execute both
$\lb \Pscr, G_1\rb$ and   $\lb \Pscr, G_2\rb$ in sequence. Both executions must succeed for the
current execution to succeed.

\item If $G$ is $G_1\seqor G_2$, then execute both 
$\lb \Pscr, G_1\rb$ and  $\lb \Pscr, G_2\rb$ in sequence. At least one of
these executions  must succeed for the current execution to succeed.

\end{numberedlist}
\end{defn}

\noindent  
The symbols $\seqand$ and $\seqor$ provide sequential executions of instructions: they allow, respectively, for the sequential conjunctive execution of
the instructions  and the sequential disjunctive execution of 
instructions.

\section{\seqweb}\label{sec:modules}

In our context, a \seqweb\ page corresponds simply to a set of $D$-formulas or a $G$-formula  with a URL. 
The module construct $mod$ allows a URL to be associated to a set of $D$-formulas or a $G$-formula.
An example of the use of this construct is provided by the 
following ``arithmetic'' module which contains some basic 
arithmetic-handling instructions written in conventional Prolog.

\begin{exmple}
 $mod(www.dau.com/arith)$.\\
$cube(X,Y) :-$ \> \hspace{4em} $Y = X * X *X.$\\  
$fib(0,1).$ \% first Fibonacci number \\  
$fib(1,1).$ \% second Fibonacci number \\               
$fib(X,Y) :-$ \> \hspace{4em} $X>1 \land X1\ is\ X-1 \land$\\
\> \hspace{4em} $X2\ is\ X-2 \land fib(X1,Z)\land$\\
\> \hspace{4em} $fib(X2,W) \land Y\ is\ W+Z.$\\
$prime(X) :-$ \> \hspace{4em}$\ldots.$\\
\end{exmple}

Our language in Section 2 permits sequential conjunction in goals.
 This allows for  sequential interactions between the user and the system.
 An example of
 the use of this construct is provided by the 
 page which does the following sequential tasks: read a number from the
user, output
the number, and then output its fibonacci number:

\begin{exmple}
 $mod(www.dau.com/query1)$.\\
$read(N)\ \seqand$ \\
 $write(N)\ \seqand$  \\
$fib(N,O)\ \seqand$\\
$write('fib:')\ \seqand$ \\
$write(O)$\\
\end{exmple}
\noindent 
 These  pages can be made available in specific contexts by explicitly
mentioning the URL via a hyperlink. For example, consider a goal 
$www.dau.com/arith \Ra  www.dau.com/query1$.
Solving this goal  has the effect of executing $query1$ 
 with respect to the instructions in  $arith$.
 
Imagine the number typed in is twin prime in the sense that both the number and
its successor by two are  prime. In  such cases we want to
indicate in addition that the number is twin 
prime.
One way of displaying this extra information is through the sequential
disjunctive goals. 
 The following is a modification of the
$query1$ page based on this idea.

\begin{exmple}
 $mod(www.dau.com/query2)$.\\
$read(N)\ \seqand $\\
$write(N)\ \seqand$  \\
$($ \\
\>\>  $(fib(N,O)\ \seqand$\\
\>\>  $ write('its\ fib:')\ \seqand$ \\
\>\>   $write(O))$ \\
\>$ \seqor$ \\
\>\>  $((prime(N) \land prime(N+2))\ \seqand $\\
\>\>  $write('twin\ prime'))$\\
 $)$ \\
\end{exmple}
\noindent 
 It is  
interesting to note that  the above task requires the parallel-AND,
the sequential-AND, and the sequential-OR executions of instructions altogether. 

Consider the task of measuring the blood pressure of a patient, \ie, 
$bp(X,Y)$, where $bp(X,Y)$ means
``patient $X$ has blood pressure $Y$''.  Successfully solving this task 
 requires performing the measurement by a nurse dynamically and, in addition,
 the outcome to be less than 180.
The following is a program based on this idea.

\begin{exmple}
 $mod(www.dau.com/nurse)$. \% DongA hospital\\
$bp(X,Y) :-$ \> \hspace{4em} $read(Y) \land (Y < 180).$\\
\end{exmple}

Now imagine a system that diagnoses the blood pressure of a patient by having him taken the
measurements twice in sequence.
One way of expressing this system is the following query.
 
\begin{exmple}
 $mod(www.dau.com/query3)$.\\
$read(X)\
 \seqand $ \% $X$ is a patient\\
 $(((bp(X,Y1)\seqand  write(Y1))\ \seqor$  $(bp(X,Y2)\seqand  write(Y2)))$\\
\end{exmple}
\noindent  Consider a goal $www.dau.com/nurse$ $\Ra$  $www.dau.com/query3$.
Solving this goal  has the effect of executing $query3$ 
 with respect to the instructions in  $nurse$.
At the beginning, the system waits until the user specifies a patient $X$. After $X = Kim$ is selected, the system
 requests the nurse to measure Kim's blood pressure. If this is a success, the nurse reports the outcome to the system.
Once the outcome  -- say $Y1 = 150$ -- is received,  the system requests the nurse to measure Kim's blood pressure
once more. Again once the outcome  -- say $Y2 = 130$ -- is received, the system reports $Z = 130$.

\section{Conclusion}\label{sec:conc}
\newcommand{\sqc}{\mbox{\small \raisebox{0.0cm}{$\bigtriangleup$}}} % sequential conjunction
\newcommand{\sqd}{\mbox{\small \raisebox{0.049cm}{$\bigtriangledown$}}} % sequential disjunction

In this paper, we have considered an extension to LogicWeb with  
sequential hereditary Harrop formulas. This extension allows goals of 
the form $G\seqand G$ and $G\seqor G$ 
 where $G$ is a goal. 
These goals are 
 particularly useful for sequential executions of instructions.

 In \cite{Jap08}, Japaridze introduced two sequential operators $\sqc$ and $\sqd$.
These are additive versions of our $\seqand,\seqor$: For goals $A\sqc B$ or $A\sqd B$ to succeed, 
it is required to succeed in only one of the tasks.
 These features in goal formulas are probably 
desirable in terms of expressiveness.
Our future  interest is in a  language which extends sequential hereditary 
Harrop formulas with these features.

\begin{acknowledgements}
This work  was supported by Dong-A University Research Fund. 
\end{acknowledgements}



\end{document}